\documentclass[conf]{new-aiaa}
\usepackage[utf8]{inputenc}

\usepackage{graphicx}
\usepackage{amsmath}
\usepackage[version=4]{mhchem}
\usepackage{siunitx}
\usepackage{longtable,tabularx}
\usepackage{subcaption}

\usepackage{todonotes}
\usepackage{booktabs}
\usepackage{multirow}
\usepackage{colortbl}

\title{Towards a Machine Learning-Based Approach to Predict Space Object Density Distributions}

\author{Victor Rodriguez-Fernandez\footnote{Associate Professor, Department of Computer Systems Engineering, \href{mailto:victor.rfernandez@upm.es}{victor.rfernandez@upm.es}}}
\affil{Universidad Politécnica de Madrid, Madrid 28038, Spain}

\author{Sumiyajav Sarangerel\footnote{B.S. in Computer Science and Engineering, \href{mailto:sschan01@mit.edu}{sschan01@mit.edu}}, Peng Mun Siew\footnote{Research Scientist, Department of Aeronautics and Astronautics, \href{mailto:siewpm@mit.edu}{siewpm@mit.edu}}}
\affil{Massachusetts Institute of Technology, Cambridge, Massachusetts 02139, USA}

\author{Pablo Machuca\footnote{Visiting Assistant Professor, Department of Aerospace Engineering, \href{mailto:pmachuca@sdsu.edu}{pmachuca@sdsu.edu}}} 
\affil{San Diego State University, San Diego, California 92182, USA}

\author{Daniel Jang\footnote{Ph.D. Candidate, Department of Aeronautics and Astronautics, \href{mailto:djang@mit.edu}{djang@mit.edu}.}, and Richard Linares.\footnote{Rockwell International Career Development Professor and Associate Professor, Department of Aeronautics and Astronautics, \href{mailto:linaresr@mit.edu}{linaresr@mit.edu} Senior Member AIAA.}}
\affil{Massachusetts Institute of Technology, Cambridge, Massachusetts 02139, USA}

\begin{document}

\maketitle

\begin{abstract}
With the rapid increase in the number of Anthropogenic Space Objects (ASOs), Low Earth Orbit (LEO) is facing significant congestion, thereby posing challenges to space operators and risking the viability of the space environment for varied uses. Current models for examining this evolution, while detailed, are computationally demanding. To address these issues, we propose a novel machine learning-based model, as an extension of the MIT Orbital Capacity Tool (MOCAT). This advanced model is designed to accelerate the propagation of ASO density distributions, and it is trained on hundreds of simulations generated by an established and accurate model of the space environment evolution. We study how different deep learning-based solutions can potentially be good candidates for ASO propagation and manage the high-dimensionality of the data. To assess the model's capabilities, we conduct experiments in long term forecasting scenarios (around 100 years), analyze how and why the performance degrades over time, and discuss potential solutions to make this solution better.
\end{abstract}

\section{Introduction}
\lettrine{I}{n} recent years, the Low Earth Orbit (LEO) has been experiencing significant congestion due to the rapid increase in the number of Anthropogenic Space Objects (ASOs) such as satellites and space debris. This escalating trend is projected to continue as multiple companies, including SpaceX, Amazon, and Astra Space, plan to launch large constellations of hundreds to thousands of satellites. The resulting dense and complex operating environment elevates the risk of collisions and debris generation, posing substantial challenges for space operators. Not only does this situation threaten the safety of flight and mission success in the short run, but it also jeopardizes the long-term viability of the LEO environment for scientific, commercial, and national security uses. Hence, understanding and modeling the evolution of the space environment is crucial for ensuring its sustainability and informing strategies for effective space traffic management. 

A variety of models have emerged to examine this evolution and calculate the orbital capacity, which is referred to as the number of satellites that can feasibly be situated in LEO \cite{doi:10.2514/1.G007208}. Traditional proprietary models, developed by organizations like NASA's LEGEND \cite{Liou2004}, ESA's DELTA \cite{Martin2004}, JAXA's IMPACT \cite{SORGE201640} among others \cite{Dolado-perez2013, Wang2019}, have been complemented by newer open-source initiatives such as the MIT Orbital Capacity Tool (MOCAT) and its various versions \cite{Jang2023, doi:10.2514/1.A35579}. Most of these models operate by propagating all the ASOs forward in time, utilizing physical models of spacecraft dynamics. This methodology incorporates factors such as atmospheric drag, solar radiation pressure, third-body perturbations, and space weather, in addition to simulated collisions and explosions. Although this technique provides detailed information about individual ASOs, it is computationally demanding and time-intensive. There are also more recent approaches that propose a Multi-layer Temporal Network Model (MTN) to explore space environment dynamics through network interactions among space objects, offering a different perspective on modeling and analyzing the complex dynamics of space environment evolution \cite{wang2023multi}.

\begin{figure}[hbt!]
\centering
\includegraphics[width=\textwidth]{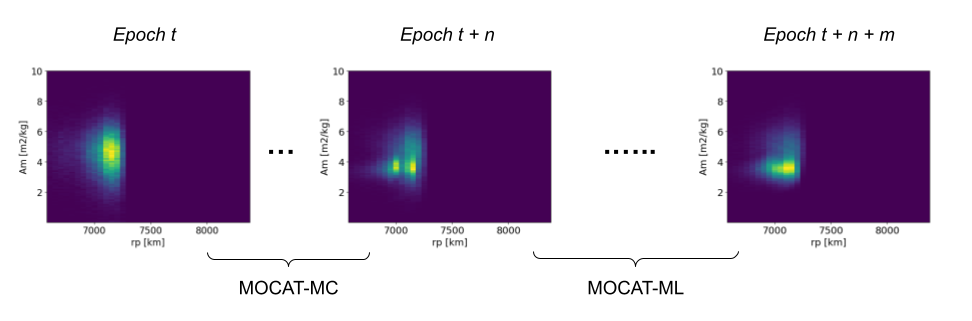}
\caption{MOCAT-ML is designed to propagate phase space density distributions up to epoch $t + n + m$, given a preliminary sequence of distributions up to $t + n$, which are computed by a reliable and accurate method such as MOCAT-MC.}
\label{fig:general}
\end{figure}

To address the inherent computational efficiency challenge in these methods, we suggest a machine learning-based strategy to study the evolution of the space environment. Our framework, named MOCAT-ML, has been created to extend the capabilities of the MIT Orbital Capacity Tool. In this work, we show the current development of a predictive model in MOCAT-ML, specifically aimed at forecasting ASO density distributions over the phase space of orbital elements and the area-to-mass ratio $A/m$. It does this by utilizing a dataset of simulations of the evolution of ASOs in different scenarios and initial conditions, generated by the validated and precise Monte Carlo variant of MOCAT, MOCAT-MC \cite{Jang2023}. Once trained and validated, the primary purpose of the new model is to accelerate the propagation of the density distribution of ASOs for new scenarios. Various data representations, neural architectures, and training strategies are examined and compared with respect to efficiency and accuracy, spanning from autoencoders to recurrent neural networks and transformers. As depicted in Fig. \ref{fig:general}, this new model isn't intended to replace other models, but to be utilized in a cooperative manner, where the initial epochs of the evolution are provided by a reliable method (e.g. MOCAT-MC), and then the machine learning model completes the rest of the propagation.

The rest of the paper is structured as follows: Section \ref{sec:backgrounds} provides a comprehensive overview of the existing literature and foundational concepts in both the MIT Orbital Capacity Tool (MOCAT) and deep learning, the main technology utilized in the paper. Section \ref{sec:dataset} details the dataset utilized in our study, emphasizing its significance and sources. In Section \ref{sec:approaches}, we introduce our novel deep learning approaches designed to support the evolution of the space environment through the prediction of phase space density distributions. Section \ref{sec:experimentation} presents a thorough examination of the experiments conducted, including setup, implementation details, and preliminary results, benchmarking the proposed approaches and testing the best of them for long-term forecasts. Section \ref{sec:perspectives}  discusses the implications of our findings and potential future research directions. Finally, in Section \ref{sec:conclusions} we summarize the key insights and contributions of our study.

\section{Backgrounds}
\label{sec:backgrounds}

\subsection{MIT Orbital Capacity Tool}
The MIT Orbital Capacity Analysis Tool (MOCAT) is a set of modeling tools created to understand the long-term sustainability of the LEO environment. There are two distinct methods -- the Source-Sink Evolutionary Model (SSEM) and the Monte-Carlo (MC) methods. MOCAT-SSEM \cite{MOCATSSEM2023} divides the entire LEO population into a few object types, which interact at the population level within each altitude shell.  The dynamics that govern the population is represented by a set of differential equations, which makes this method computationally efficient. MOCAT-MC \cite{Jang2023} is a Monte-Carlo tool that propagates all LEO objects and models interaction at each timestep. As an MC-based tool, MOCAT-MC simulates the evolution of the LEO environment and models the interaction of each object with one another to understand the population evolution. The Cube method is used for stochastically sample collisions, and the NASA Standard Breakup model is used for collision debris generation, while a variety of atmospheric model and propagators can be used. For this work, MOCAT-MC is used to create a time-series set of propagated orbital objects.

\subsection{Deep learning concepts relevant to this work}
Since the inception of neural networks, both network architecture and training algorithms have undergone numerous iterations, resulting in substantial advancements in performance, training efficiency, and data utilization. Meanwhile, some architectures have been crafted to specialize in handling specific data types or tasks, leveraging spatial and/or temporal relationships within the data. Examples of such architectures include Convolutional Neural Networks (CNNs) \cite{lecun1989generalization}, Recurrent Neural Networks (RNNs) \cite{medsker1999recurrent}, and Transformer networks \cite{vaswani2017attention}. Deep learning is a natural extension of neural networks, where multiple neural networks are connected in sequence or in parallel to form deep architectures. These deep architectures can learn hierarchical representations of data, with each layer capturing abstractions of the data at different ``depth'' levels. CNNs are a class of deep neural networks specially designed to handle grid-structured data. CNNs utilize small filters to slide or convolve across the input data, capturing local patterns and features efficiently. On the other hand, RNNs are designed to handle temporal data, where they process sequential data by maintaining a hidden state that is updated at each time step. This mechanism allows RNNs to capture temporal dependencies within the data. Variations of RNN architectures include Long Short-Term Memory (LSTM) Networks \cite{hochreiter1997long} and Gated Recurrent Units (GRUs)  \cite{gers2000learning}. Transformer networks are also designed to handle temporal data, but in a more efficient fashion compared to traditional RNNs. Transformer networks utilize self-attention mechanisms to weigh the importance of different data points in a sequence, enabling them to capture dependencies regardless of the distance between data points. Unlike traditional RNNs, Transformer networks process the entire sequence of data in parallel, further enhancing their computational efficiency.
In this work, we also utilize an autoencoder, a type of unsupervised machine learning \cite{hinton2006reducing}. An autoencoder consists of an encoder and decoder pair, with the primary objective of learning an efficient encoding of the input data. The model aims to facilitate the reconstruction of the input data from compressed encoding.

\section{Dataset}
\label{sec:dataset}
As aforementioned, MOCAT-MC is used to propagate both an initial population and a yearly number of newly launched satellites along time, considering interactions between objects (i.e., collisions) that result in added debris objects to the Earth-orbit environment, and natural dynamics such as those induced by Earth's oblateness ($J_2$ effect) and atmosphere, which result in changes in each object's orbital elements and potential decay into the atmosphere. The initial population is based on two-line element sets (TLEs) corresponding to January 2023, and the yearly launch rate repeats the frequency and altitude pattern of launches in the 2018--2022 period. Both the initial population and number of yearly launches can be multiplied by arbitrary factors to simulate alternative scenarios. Here, both the initial population and yearly launches are multiplied by a factor of $15$, which reduces the sparsity of interactions between objects.

MOCAT-MC considers various stochastic processes such as probability of collision or number and physical parameters (e.g., mass and size) of newly generated debris objects. Multiple (Monte Carlo) simulations are then performed to gain insight into the expected time evolution of a given scenario. Ultimately, for each simulation of a scenario, MOCAT-MC provides a sequence of each object's state (e.g., position, velocity, or orbital elements along time).

Following a similar representation to that in \cite{Giudici2023}, at each time step, we can process MOCAT-MC results to generate density distributions of objects in various phase spaces: e.g., area-to-mass ratio of objects, $A/m$, \textit{versus} their perigee, $r_p$ (as in Fig. \ref{fig:general}), inclination \textit{versus} apogee, etc., providing a graphical and intuitive representation of number of objects within certain phase-space regions, along time. This work primarily focuses on predictions of the density distribution over the area-to-mass ratio and perigee phase space: $A/m$ is discretized into 99 evenly distributed bins in a log-scale (from $10^{-5}$ $m^2/kg$ to $10^1$ $m^2/kg$, which is comprehensive of all objects in the MOCAT-MC simulations), and $r_p$ is discretized into 36 bins of 50 km in height, from 200 km to 2,000 km altitudes (objects with a perigee below 200 km are considered to have decayed into the atmosphere, and objects above 2,000 km are considered to be outside the LEO region and are not simulated).

Conclusively, the dataset employed in this work consists of multiple Monte Carlo simulations ($100$) of the same scenario (initial population and launch rate multiplied by a factor of $15$), where, at each time step within each simulation, the number of objects within each two-dimensional bin of area-to-mass ratio and perigee is known.

\section{Proposed approaches}
\label{sec:approaches}
We frame the problem of forecasting phase space density distributions from a pure data-driven perspective. No expert knowledge is included in any of the models at any part of the stage. The common factor among the three approaches proposed in this section is the use of deep learning. Each of the approaches explore different representation learning techniques and training strategies that range from classic architectures for dealing with sequential data such as the LSTM or the Transformer, to a combination of convolutional and recurrent layers that exploit the spatio-temporal nature of the data.

\subsection{Decoupled one-dimensional autoencoder and TST forecaster (\texttt{Decoupled 1D AE + TST})}
Based on the expertise of the authors in deep learning applied to time series in the space domain \cite{STEVENSON20232660, briden_transformer-based_2023}, we decided to start tackling this problem from a time series forecasting point of view. However, as seen in section \ref{sec:dataset}, the data used in this problem is not a one-dimensional time series, but a sequence of two-dimensional density distributions. Therefore, we have to encode first the density distributions into one-dimensional vectors. 

\begin{figure}
    \centering
    \includegraphics[width=\linewidth]{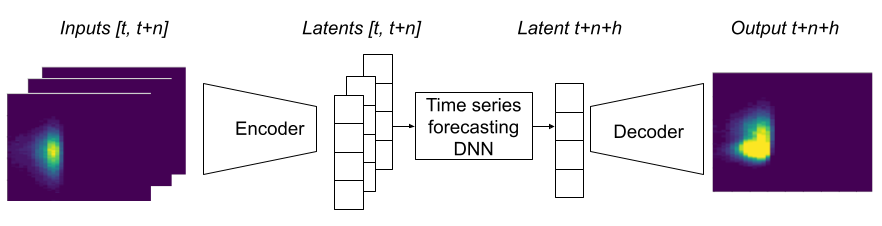}
    \caption{Encoder-decoder architecture to lower the dimensionality of the input space before applying a Deep Neural Network (DNN) suitable for time series forecasting.}
    \label{fig:decoupled}
\end{figure}

As a result, we designed an encoder-decoder architecture, which can be seen in Fig. \ref{fig:decoupled}, where we first encode the sequence of input density distributions into a lower one-dimensional latent space. Next, a time series forecasting deep neural network is employed to propagate that latent state into the future. Finally, a decoder reverts the future states into the corresponding distributions in the original space of density distributions.

We chose a Transformer architecture for time series forecasting, also known as TST \cite{Yuqietal-2023-PatchTST} to be used as the forecaster component. This work is important for time series forecasting as it introduces a novel, efficient design for Transformer-based models, leveraging techniques like patching and channel independence, which have successfully been applied in other fields such as computer vision. These innovations enhance the model's ability to perform long-term forecasting with higher accuracy and efficiency compared to state-of-the-art models.

As for the encoder-decoder structure itself, we designed a simple autoencoder that processes one density distribution at a time. The encoder compresses each two-dimensional density distribution using a sequence of convolutional layers, while the decoder employs transposed convolutional (or devconvolutional) layers \cite{long2015fully} for upscaling, effectively reconstructing the original two-dimensional distributions from the latent space representation.

The reason why we call this section ``Decoupled" is that we did not train the two main components of the architecture, namely the autoencoder and the forecaster, at the same time. Instead, we trained each component separately in its own training process, keeping the validation set in common between them so that there is no data leakage from one to another. This approach is simpler and worked effectively in similar works where we tried to forecast the atmospheric density following the same strategy \cite{briden_transformer-based_2023}.

\subsection{End to end one-dimensional autoencoder and LSTM forecaster (e2e 1D AE + LSTM)}
In this network, a batch of image sequence serves as the input, traversing through an encoder layer, LSTM, and decoder layer. Each encoder and decoder layer is composed of both convolutional and linear components. The convolutional layers incorporate three Con2d layers, facilitating upsampling of the input. Subsequently, the output undergoes flattening through the subsequent linear layer, effectively downsampling the input into the encoding dimension. In this setup, the encoder processes a batch of image sequences, treating the input as a matrix with dimensions of batch size x sequence length. It then encodes this input into a matrix with dimensions of batch size x sequence length x encoding dimension. This process ensures that the encoder doesn't engage with time series data directly; instead, its role is to transform the input into vectors. Next, LSTM with hidden size 500 is used to capture the time series part of the data. Its output is in the shape of batch size x horizon x encoding dimension. Then, it runs through linear and convolutional parts of the decoder layer. Every convolutional layer has batchnorm layer and elu activation function. Dropout layers with p = 0.5 are used in the linear layers of encoder and decoder layer of the architecture. During the training, we observed a lot of instability and used gradient clipping to ensure a stable training process. Furthermore, we did not use one cycle policy which resulted in slower convergence. Instead, training with constant learning rate that was found from lr-finder produced much quicker convergence. 

\subsection{End to end bi-dimensional autoencoder and forecaster}
In our last proposed model, we integrate the strengths of convolutional and recurrent neural networks to process spatiotemporal data, keeping the latent space that goes from the encoder to the decoder as bi-dimensional feature maps. The input for this architecture is a sequence of density distributions of length $l$, and, unlike the two approaches presented above, the output in this case will not be a single density distribution but a sequence of the same length. In forecasting words, we are setting a horizon for the forecasts equal to the lookback period.  We leave for future work the design of a more flexible architecture like this that allows the use of different values for the lookback and horizons.

\begin{figure}
    \centering
    \includegraphics[width=\linewidth]{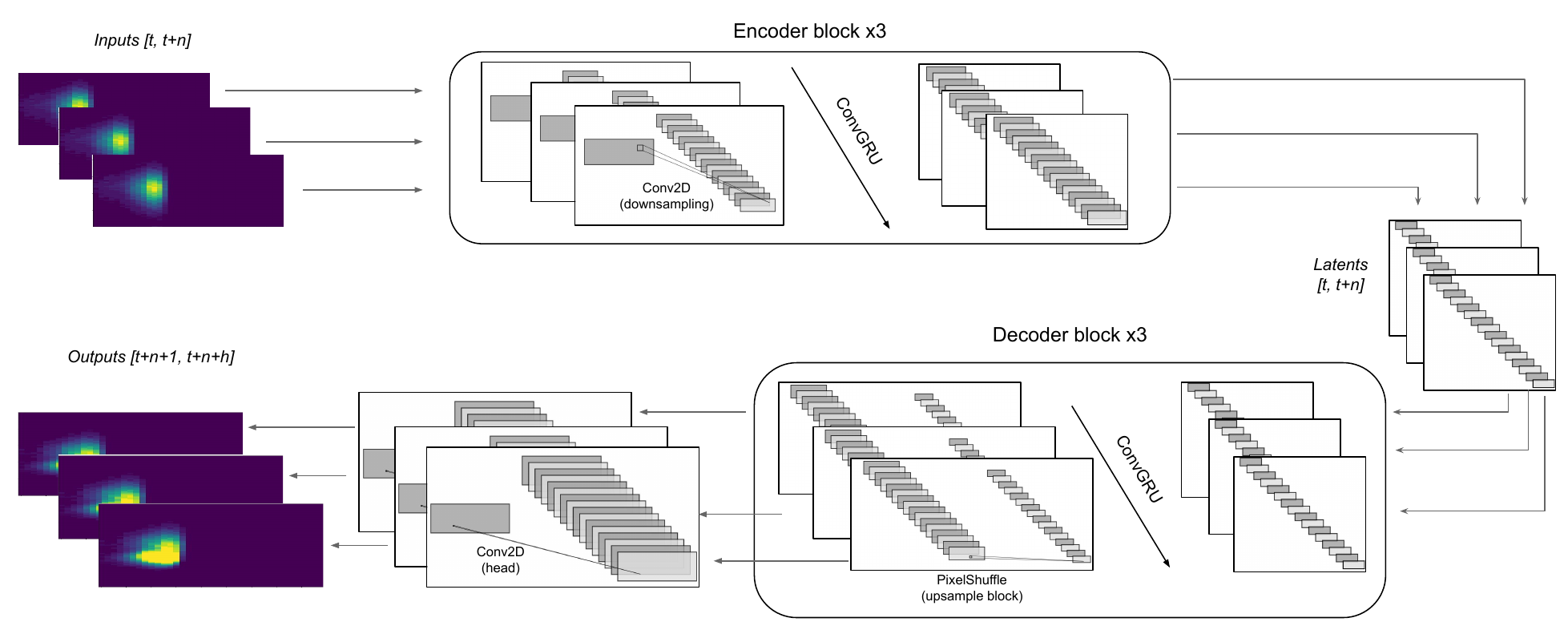}
    \caption{Overview of the encoder-decoder ConvGRU neural network. Encoder blocks (three in this example) downsample and encode temporal features, which are then upsampled and decoded by decoder blocks (also three) to predict the output sequence. Temporal coherence is maintained by ConvGRU cells, while spatial resolution is restored via PixelShuffle upsampling.}
    \label{fig:convgru}
\end{figure}

A graphical view of the architecture can be found in Fig. \ref{fig:convgru}. The architecture, again, adopts an encoder-decoder structure, where both components contain, as a key layer, the so-called \textit{ConvGRU} cells. Each ConvGRU cell in the encoder processes the spatial data across time, maintaining the spatial structure while capturing temporal dependencies. This is crucial for understanding changes in the density distributions over time. 

In the encoder design, the flow of data through convolutional layers and ConvGRU cells is carefullyw orchestrated to capture both spatial and temporal aspects of the input densities. The encoder processes the input data in several stages, with each stage consisting of a convolutional operation followed by a recurrent (ConvGRU) processing step. Initially, the input sequence is optionally passed through a \textit{CoordConv} \cite{liu2018intriguing} layer, in case spatial context wants to be emphasized. Following this, the sequence enters the first stage of the encoder, where a convolutional layer processes each density of the sequence independently. This step ensures consistent extraction of spatial features across all time steps, maintaining the integrity of spatial information in each density. Immediately after the convolutional processing, the sequence is passed to a ConvGRU cell. Unlike the preceding convolutional operation, the ConvGRU cell does not treat each density in isolation. Instead, it integrates information across the sequence, capturing how the spatial features evolve over time. This recurrent processing is crucial for understanding temporal dynamics, allowing the model to track changes and patterns that unfold through the sequence. This process of alternating between convolutional layers and ConvGRU cells continues through the encoder for several stages (three by default), progressively refining the representation of the encoded data.

The decoder, on the other hand, reconstructs the forecasted sequence from the encoded representations. At each stage, a ConvGRU cell is applied first, focusing on integrating temporal information across the encoded densities. Then, the sequence is passed through an upsample block. Each upsample block is applied independently over the time dimension, and utilizes \textit{PixelShuffle} layers, introduced in \cite{shi2016real}, designed to avoid common issues like ``checkerboard" artifacts in upscaled images. This process of alternating between ConvGRU cells and upsample blocks is repeated across the decoder stages. 

Finally, the last layer of the architecture takes the output of the decoder and applies, independently over each time step, a $1 \times 1$ convolution to match the desired spatial dimensions (the same as the input), and the channel depth, which in our case is one, since we only process one phase space at a time, the combination of perigee and area-to-mass ratio.

Because in this architecture the horizon of the forecast is greater than one, the loss function (MSE) is aggregated across all time steps, providing a comprehensive measure of the model's performance over the entire sequence. Across the entire architecture, group normalization \cite{wu2018group} is utilized as the normalization layer, enhancing training stability and model performance without the dependencies on batch size typical of batch normalization. Indeed, we found that replacing the normalization layers to batch norm completely prevented the network from learning anything during training process. The activation function utilized was ReLU in every layer of the network.

\section{Experimentation}
\label{sec:experimentation}
This section presents the experiments done with the proposed approaches in section \ref{sec:approaches}. The experimentation is divided into two primary experiments. The first one focuses on benchmarking the effectiveness of the proposed deep learning architectures in accurately performing a simple one-step ahead prediction of the phase space density distribution. The second experiment uses the best model from the aforementioned benchmark and explor  its capability in long-term forecasting.

We ran every experiment of this study in a GPU node from the MIT Supercloud \cite{reuther2018interactive}. Each node is powered by an Intel Xeon Gold 6248 processor and features 40 CPU cores. The node is equipped with 384 GB of RAM, averaging 9 GB per core, and includes two Nvidia Volta V100 GPUs, each with 32 GB of RAM. We implemented the code using the libraries \texttt{fastai} \cite{howard2020deep} and \texttt{tsai} \cite{tsai}, which are both written on top of Pytorch. Our source code has been made publicly available on Github \footnote{\url{https://github.com/ARCLab-MIT/mocat-ml}}  

\subsection{Benchmark of proposed approaches}
\label{sec:benchmark}

In this section, we perform a benchmark among the approaches proposed in section \ref{sec:approaches}, under the same data and conditions. We utilize a dataset comprising $100$ simulations from MOCAT-MC. For each of the simulations, we take the sequence of density distributions over the phase space of the perigee ($r_p$) and the area-to-mass ratio ($A_m$). Therefore, our data can be represented as a four-dimensional array with dimensions corresponding to the number of simulations, time steps, and a discretized density distribution grid defined by area-to-mass ratio (height of the grid) and perigee (width of the grid). Each time step in our simulations represents a two-week interval. Since we are only interested in a side-by-side comparison in this experiment, we only get the first 2 years of data of each simulation (i.e., around 100 distributions per sequence).

To prepare this dataset for input into our approaches, we employ a sliding window process. This process creates overlapping windows across the time dimension of each simulation. Each window encompasses a specific number of consecutive time steps, defined by the sum of the \textit{lookback} and the \textit{horizon}. The lookback parameter determines the number of past time steps included in each window, capturing the historical data necessary for the model to learn temporal patterns. The horizon parameter represents the number of future time steps the model aims to predict. The choice of lookback and horizon values is critical, as it balances the amount of historical information the model can access against its predictive capacity for future states. For this benchmark, we keep the horizon fixed as one (i.e., the task will be a next-step prediction), while we do multiple training runs with increasing values of the lookback, ranging from 2 steps to 16. The stride to perform the sliding window is fixed as 2 steps. Given that the bi-dimensional ConvGRU approach works only in a setup where the horizon is equal to the lookback (i.e., it forecasts the same number of densities it gets), for this experiment we train that approach restricting the loss function to be computed only in the first element of the output, to compare side by side the performance in just next-step prediction.

\begin{table}[ht]
    \centering
    \caption{Configuration for each of the proposed approaches on the benchmark. AE stands for AutoEncoder, LR stands for Learning Rate. The number of filters are given as an array with the value per layer.}
    \label{tab:config_table}
    \begin{subtable}[t]{0.32\textwidth}
        \centering
        \caption{Decoupled 1D AE + TST}
        \label{tab:subtable1}
        \begin{tabular}{ll}
            \toprule
            Batch size & 512 \\
            Conv. kernel size & (3,3)\\
            Conv. padding & zero padding\\
            \# Filters encoder & [8, 16, 32]\\
            \# Filters decoder & [32, 16, 8]\\
            LR & 0.01\\
            N. Epochs & 20\\
            \# Parameters (AE) & 1,652,769\\
            \# Parameters (TST) & 536,896\\
            TST model type & Plus\\
            Optimizer & Adam \\
            Loss function & MSE \\
            \bottomrule
        \end{tabular}
    \end{subtable}
    \hfill 
    \begin{subtable}[t]{0.32\textwidth}
      \centering
        \caption{e2e 1D AE + LSTM}
        \label{tab:subtable2}
        \begin{tabular}{ll}
            \toprule
            Batch size & 128 \\
            Conv. kernel size & (3,3)\\
            Conv. padding & zero padding\\
            LSTM hidden size & 500\\
            Dropout rate & 0.5\\
            \# Filters encoder & [24, 48, 16]\\
            \# Filters decoder & [16, 48, 24]\\
            Norm & BatchNorm\\
            LR & 0.01\\
            \# Epochs & 20\\
            \# Parameters & 709,985\\
            Act. function & eLU \\
            Optimizer & Adam \\
            Loss function & MSE \\
            \bottomrule
        \end{tabular}
    \end{subtable}
    \hfill 
    \begin{subtable}[t]{0.32\textwidth}
        \centering
        \caption{e2e 2D ConvGRU}
        \label{tab:subtable3}
            \begin{tabular}{ll}
            \toprule 
            Batch size & 32 \\
            Conv. kernel size & (3,3) \\
            ConvGRU kernel size & (5,5) \\
            ConvGRU padding & zero padding \\
            Dropout rate & 0.2 \\
            N. Filters encoder & [16, 64, 96] \\
            N. Filters decoder & [96, 64, 16] \\
            LR schedule & 1-cycle \\
            Max. LR & 0.001 \\
            N. Epochs & 20 \\
            N. Parameters & 4,198,401 \\
            Act. function & ReLU \\
            Optimizer & Adam \\
            Loss function & MSE \\
            \bottomrule 
            \end{tabular}
    \end{subtable}
\end{table}

The individual configuration for each of the proposed approaches can be seen in Table \ref{tab:config_table}. For the optimizer, i.e., the component that updates the parameters of the neural network, we employed the popular Adam optimizer \cite{kingma2014adam} in all cases. For the learning rate, we used a dynamic learning rate with a ``1cycle” policy \cite{smith2017cyclical}, which schedules the learning rate and momentum with cosine annealing. The maximum learning rate is set as $1e^{-3}$ in all cases too. As for the loss function, we employ the Mean Squared Error (MSE), that averages the difference between the predicted number of ASOs and the real value for every cell of the density distribution.   

\begin{table}[ht]
\centering
\caption{Comparison of the different methods for varying lookback values, in terms of the average validation loss at the end of training (MSE) and the inference time per sample. Cells with best values are highlighted.}
\label{tab:benchmark_results}
\resizebox{\textwidth}{!}{
\begin{tabular}{|c|cc|cc|cc|cc|cc|}
\hline
\multirow{2}{*}{L} & \multicolumn{2}{c|}{Approach a (d=64)} & \multicolumn{2}{c|}{Approach a (d=128)} & \multicolumn{2}{c|}{Approach b (d=64)} & \multicolumn{2}{c|}{Approach b (d=128)} & \multicolumn{2}{c|}{Approach d} \\
 & MSE & Time (ms) & MSE & Time (ms) & MSE & Time (ms) & MSE & Time (ms) & MSE & Time (ms) \\ \hline
1 & 1276.4k±308.3k & 3.38±0.17 & 595.1k±124.6k & 3.46±0.08 & 589.27±51.31 & \cellcolor{gray!25}\textbf{1.48±0.52} & 583.88±45.86& 1.79±0.59 & \cellcolor{gray!25}47.37±22.66 & 24.95±20.35 \\
2 & 1252.5k±474.8k & 3.72±0.18 & 652.8k±115.5k& 3.76±0.12 & 588.48±43.45 & \cellcolor{gray!25}1.52±0.50 & 581.85±39.60 & 1.55±0.54& \cellcolor{gray!25}49.12±19.10 & 17.66±0.25 \\
4 & 488.8k±238.8k & 3.69±0.12 & 795.1k±586.0k & 3.76±0.12 & 597.49±42.94 & \cellcolor{gray!25}1.58±0.47 & 570.94±31.22 &  1.61±0.48 & \cellcolor{gray!25}\textbf{29.53±8.23} & 27.50±0.72 \\
8 & 690.1k±375.0k & 3.72±0.11 & 340.1k±133.1k & 3.74±0.08 &  493.28±32.89 & \cellcolor{gray!25}1.80±0.51 & 
540.87±37.19& 1.80±0.55 & \cellcolor{gray!25}40.98±6.54 & 45.56±0.55 \\
16 & 262.1k±169.2k & 3.78±0.19 & 89.3k±25.6k & 3.75±0.38 & 462.98±16.68 & \cellcolor{gray!25}2.80±0.52 & 
453.70±18.00 & 2.80±0.53 & \cellcolor{gray!25}45.23±8.28 & 84.77±0.19 \\ \hline
\end{tabular}
}
\end{table}

The comparison among approaches was done in terms of the value of the validation loss after the last training epoch (the lower the better). The validation set was obtained by holding out randomly all the samples for 20\% of the simulations (i.e., 20 in this case). Additionally, we also measure the inference time per sample to have a sense of how fast this solution could be, which, at the end, it is the main benefit of this whole work over the use of just a Monte Carlo solution like the one provided by MOCAT-MC. The results are shown in Table \ref{tab:benchmark_results}. As it can be seen, there are orders of magnitude of difference in the MSE among the three approaches, where the decoupled approach performs the worst and the end-to-end 2D ConvGRU performing best. It is noteworthy that, even when using a sophisticated and state of the art Transformer architecture in the forecaster module of the decoupled approach, the fact that the training was not a single back-propagation but two separated ones is a key element to failure, regardless of the accuracy of each component and of the dimensionality of the latent space.

We also see that the evolution of the MSE with respect to the lookback does not follow a clear decreasing trend as one might expect. In fact, if we focus on the best performing approach (last column of the table), we see that the benefit of using more than just the last density to predict the next one is not aligned with a meaningful decrease of the loss. This might be due to the low pace with which, naturally, the space environment situation changes in such a small period of time as two weeks (one step in the dataset). On the other hand, we find a sweet spot at $L=4$ for approach that achieves the minimum loss of the entire table. We argue that this is just a matter of the configuration of the benchmark, and that, given enough time to train, the configurations with longer horizons would lead to the best performances.

\begin{figure}[ht!]
    \begin{subfigure}[b]{0.3\textwidth}
        \centering
        \includegraphics[width=\linewidth]{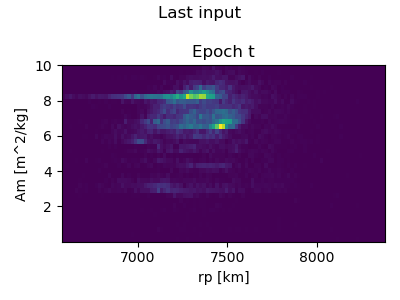}
        \includegraphics[width=\linewidth]{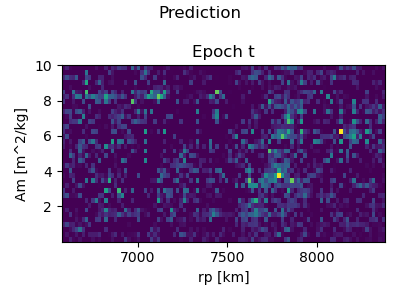}
        \includegraphics[width=\linewidth]{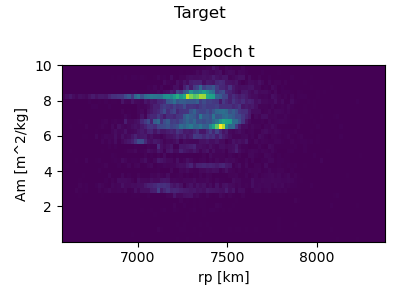}
        \caption{Decoupled 1D AE + TST}
    \end{subfigure}
    \hfill 
    \begin{subfigure}[b]{0.3\textwidth}
        \centering
        \includegraphics[width=\linewidth]{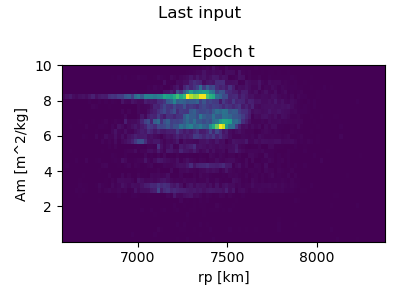}
        \includegraphics[width=\linewidth]{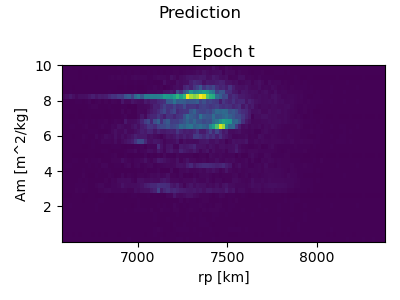}
        \includegraphics[width=\linewidth]{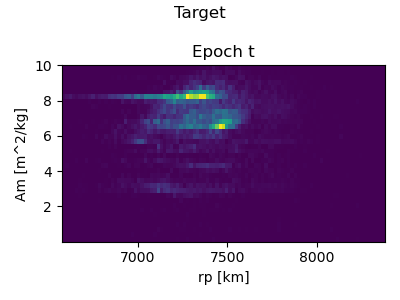}
        \caption{e2e 1D AE + LSTM}
    \end{subfigure}
    \hfill 
    \begin{subfigure}[b]{0.3\textwidth}
        \centering
        \includegraphics[width=\linewidth]{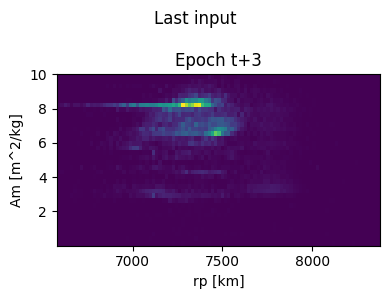}
        \includegraphics[width=\linewidth]{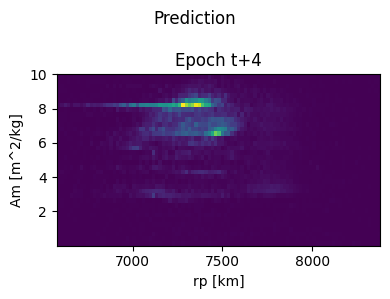}
        \includegraphics[width=\linewidth]{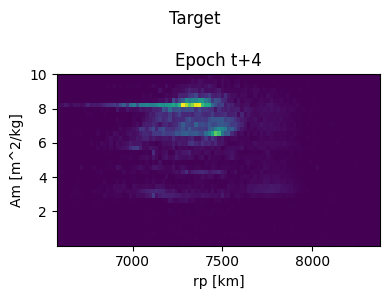}
        \caption{e2e 2D ConvGRU}
    \end{subfigure}
    \caption{Comparison of the predictions (along with last input of the lookback sequence) of the sample with worst validation loss among the different proposed approaches}
    \label{fig:top_losses}
\end{figure}

To get a visual grasp on the differences in the quality of each of the proposed approaches, we show in Fig. \ref{fig:top_losses} the samples of the validation set that got the largest validation loss (i.e., the worst predictions), for the models trained with lookback 4. As it can be seen, we can confirm here the poor performance of the decoupled approach, which makes it valueless even for such a simple benchmark, where, at the end, just using an identity function would be a strong baseline. On the other hand, both the e2e 1D AE+LSTM and the e2e 2D ConvGRU approaches seem to respect the shape of the input and create, at least, faithful reproductions of it, even if we saw in Table \ref{tab:benchmark_results} that they were far from each other in terms of MSE. 

\subsection{Long term prediction}
In this section, we select the best model from the previous benchmark and train it again on the full dataset, i.e., on a dataset of 100 simulations with a total of 2435 epochs per simulation, which span a total of around 93.4 years of data (the time-step between epochs is two weeks). The selected model is the e2e 2D ConvGRU, trained with a lookback and a horizon of 4 steps (as a reminder, this approach always generates the same number of steps it is fed with). The rest of the training configuration is kept the same as in Table \ref{tab:subtable3}, with the exception of the stride, for which we employ a value of 4 this time (therefore there are no overlapping windows in the dataset). Additionally, this time the loss function (MSE) is computed for the whole forecasted density sequence, and not only for the first element of the forecast as it was done before.

\begin{figure}[ht!]
    \begin{subfigure}[b]{0.5\textwidth}
        \centering
        \includegraphics[width=\linewidth]{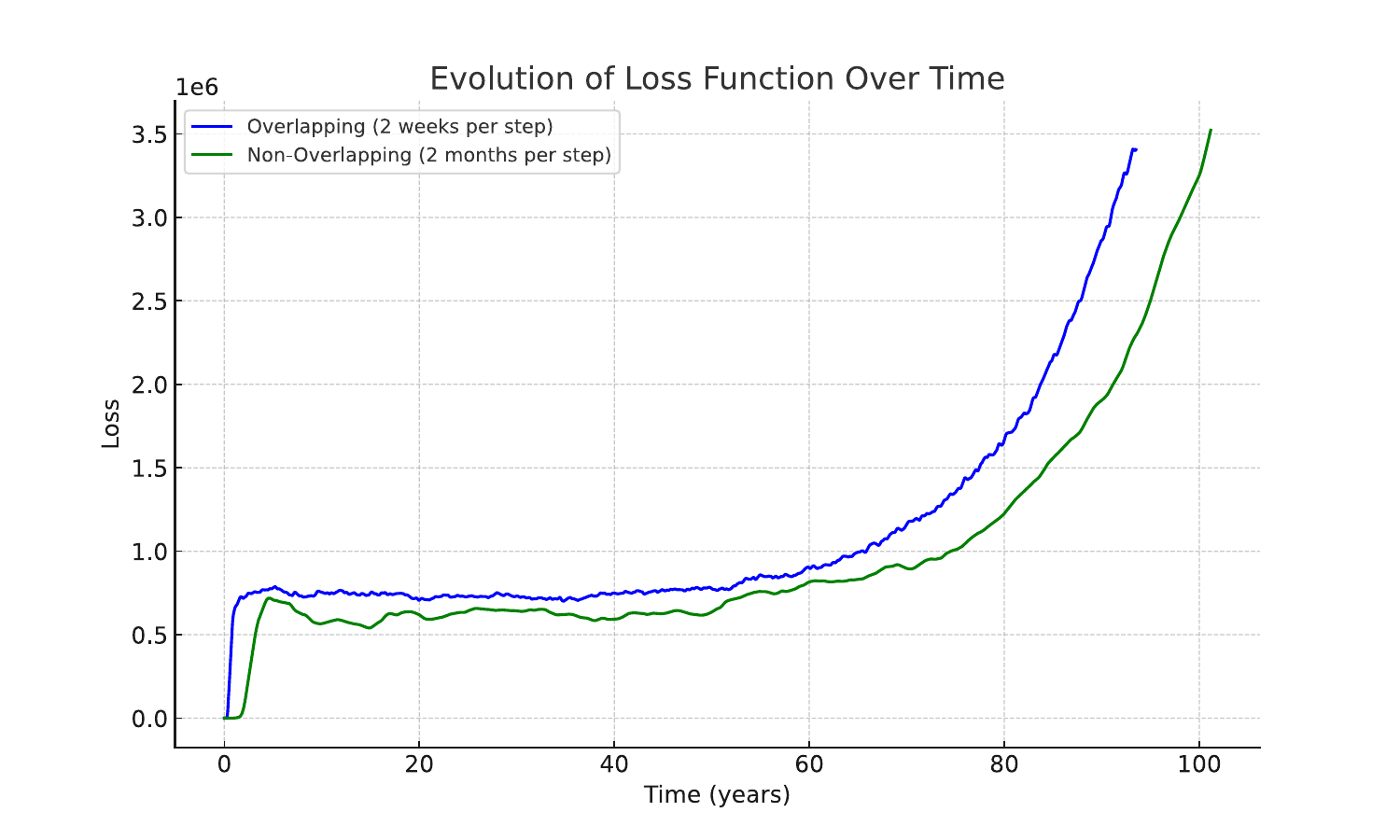}
    \end{subfigure}
    \hfill 
    \begin{subfigure}[b]{0.5\textwidth}
        \centering
        \includegraphics[width=\linewidth]{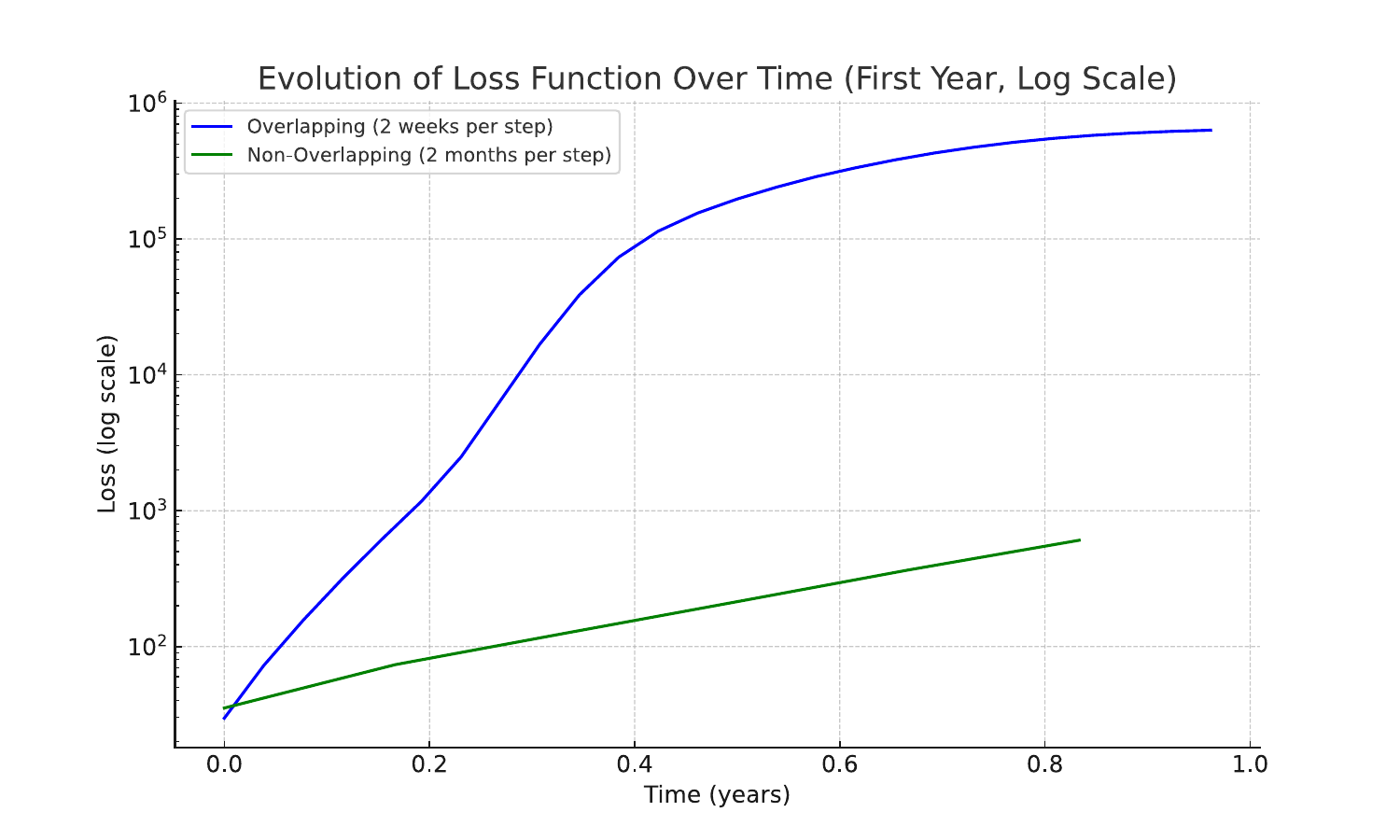}
    \end{subfigure}
    \caption{Comparative evolution of loss functions for overlapping (2 weeks per step) and non-overlapping (2 months per step) forecasting strategies: full period and first year analysis.}
    \label{fig:ltf_losses}
\end{figure}

Long term predictions are achieved by performing \textit{iterative forecasting}. This simple idea consists in calling the model many times in a loop, using in each iteration the outputs that the model produced in the previous one. 

In Fig. \ref{fig:ltf_losses}, we show the evolution of the average validation loss of the forecasts over time, for two different forecasting strategies, applied on two different trained models. The first one, that we refer to as ``non overlapping", outputs a sequence of the four predicted distributions that come directly after the input data (another sequence of four densities), i.e., it creates a two months-ahead forecast. The ``overlapping" strategy, on the other hand, produces a sequence of four density distributions too, but instead of being subsequent to the input, they actually overlap with it partially, in a way that effectively, the predicted data only spans a period- of two weeks ahead ahead of the last element of the input (i.e., it is effectively a one-step forecast as we did in the benchmark). This strategy is common in RNNs, since it allows for fine-grained one-step predictions in models that architecturally output always the same length as the input (a common pattern in seq2seq models).

On the left side of the figure, we see how both strategies fail in forecasting the full period of data (around 100 years), reaching values of the loss six orders of magnitude larger than the ones we saw in the previous benchmark (see section \ref{sec:benchmark}). In fact, only in the first year (right panel of the figure) we can appreciate a huge increase of the loss function, especially in the overlapping approach. This can be due to, despite the overlapping task being easier at each iteration (part of the prediction can be learned from the leaked input itself), it needs four times more iterations to span the same period of time than the non-overlapping approach, also consuming four times more time in the process. This leads us to conclude that, in order to explore long term forecasting strategies for this problem, we might investigate the creation of models that forecast data with ``gaps" between the input and the output, in order to reduce the number of iterations needed to span long periods of time, which is what causes a bigger degradation in this type of architectures.

\begin{figure}[ht]
    \begin{subfigure}[b]{\textwidth}
        \begin{subfigure}{.23\textwidth}
            \centering
            \includegraphics[width=\linewidth]{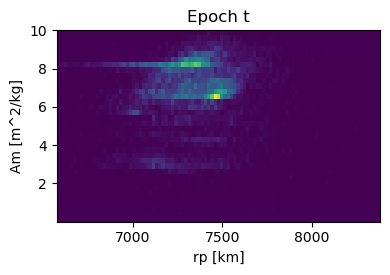}
        \end{subfigure}
        \hfill 
        \begin{subfigure}{.23\textwidth}
            \centering
            \includegraphics[width=\linewidth]{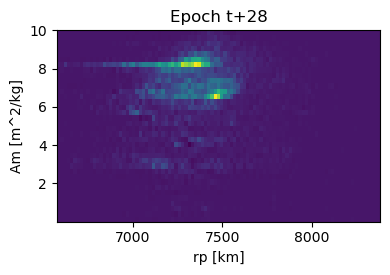}
        \end{subfigure}
        \hfill 
        \begin{subfigure}{.23\textwidth}
            \centering
            \includegraphics[width=\linewidth]{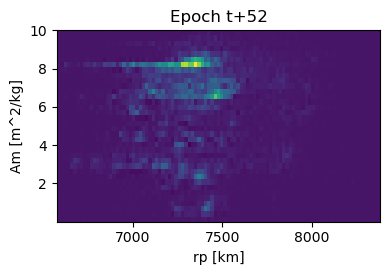}
        \end{subfigure}
        \hfill 
        \begin{subfigure}{.23\textwidth}
            \centering
            \includegraphics[width=\linewidth]{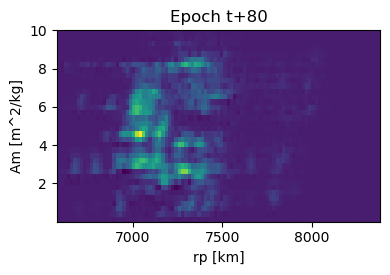}
        \end{subfigure}
        \caption{Predictions}
    \end{subfigure}
    \vspace{2em} 
    \begin{subfigure}[b]{\textwidth}
        \hspace{.23\textwidth} 
        \hfill 
        \begin{subfigure}{.23\textwidth}
            \centering
            \includegraphics[width=\linewidth]{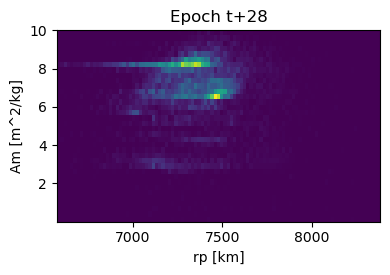}
        \end{subfigure}
        \hfill 
        \begin{subfigure}{.23\textwidth}
            \centering
            \includegraphics[width=\linewidth]{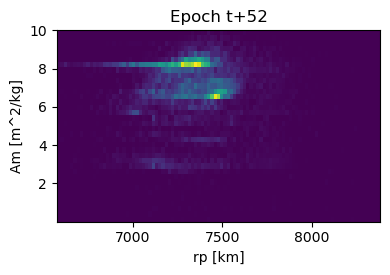}
        \end{subfigure}
        \hfill 
        \begin{subfigure}{.23\textwidth}
            \centering
            \includegraphics[width=\linewidth]{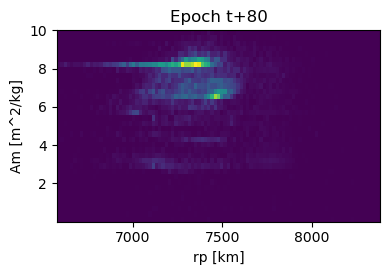}
        \end{subfigure}
        \caption{Targets (truth) for years 1 (26 epochs-ahead), 2 (52 epochs-ahead) and 3 (80-steps ahead)}
    \end{subfigure}
    \caption{Iterative yearly forecasts (top) vs true densities (bottom), produced by a ConvGRU model with a lookback and horizon of four time steps (with a step size of two weeks). Each epoch represents a two-week step}
    \label{fig:ltf_preds_vs_targs}
\end{figure}

To have a visual grasp of this degradation of performance in the models on long-term horizons, we show the predictions for one random simulation of the validation set in Fig. \ref{fig:ltf_preds_vs_targs}, and contrast them with the real densities at those epochs. As it can be seen, already in the first year, after 26 epochs (1 year, considering that each epoch spans two weeks), the background of the density map starts getting ``washed", and we can see that areas which should be constantly zero across the evolution of the space environment (e.g., edges of the density map) merge themselves, creating a lose of sharpness that is more accentuated at epoch 52 (year 2). Finally, the moment we analyze the predictions on the third year (right side of the figure), the characteristic shape of the scenario that we are using in this experimentation (See the description in Section \ref{sec:dataset}) is completely lost. This can explain why the evolution of the loss we saw in Fig. \ref{fig:ltf_losses} plateaus after the first years. The prediction becomes then a blurry distribution where the loss function is already too high to represent anything meaningful. Further study would be needed to explain the second bump in the loss in the last iterations of the curve.

\section{Perspectives}
\label{sec:perspectives}
From a machine learning point of view, there are several challenges associated with the task of propagating density distributions that we faced in this work, and that we will face in subsequent iterations of it too:
\begin{itemize}
    \item The \underline{long term horizon} of the propagation (up to a hundred years ahead).
    \item The \underline{high-dimensionality} of the problem. 
    \item The \underline{absence of real-world data and benchmarks} to assess the performance and generalization of the method.
\end{itemize}

With respect to the first challenge, we showed in our experimentation (See section \ref{sec:experimentation}) how, even a relatively sophisticated architecture such as the end-to-end 2D ConvGRU (See section \ref{sec:approaches}), which achieves decent performance in the one-step prediction baseline, fails when used iteratively to span forecasting periods of more than two years. Needless to say, there is much work to do in terms of hyperparameter tuning, design of new architectures, and implementation of more efficient forecasting strategies other than a naive iteration. However, a different perspective that we plan to implement to solve this problem is to \textit{hybridize} the Monte Carlo (MOCAT-MC)/Machine learning (MOCAT-ML) predictions. This would imply that the first $n$ epochs of the propagation are provided by MOCAT-MC and the remaining $m$ (with $m > n$) are provided by MOCAT-ML (See Fig. \ref{fig:general}) until the uncertainty of the latter is estimated to be over a tolerable threshold, where the MC approach could be used again as a way of recalibrating the process. This opens up a new research line on how to properly quantify uncertainty in this application of deep learning models.

\begin{figure}[h]
    \centering
    \includegraphics[width=\linewidth]{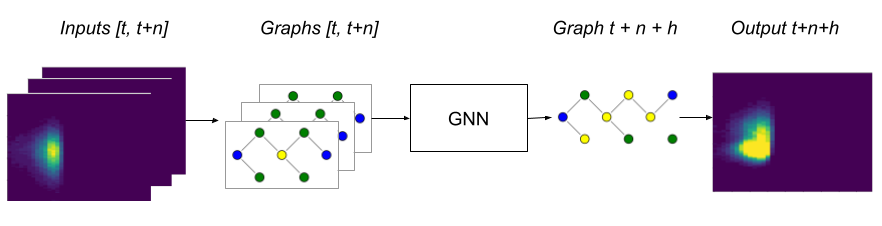}
    \caption{Perspectives on how to tackle the high-dimensionality of the problem: represent directly the input space as a graph, and train using a Graph Neural Network (GNN).}
    \label{fig:gnn}
\end{figure}

Concerning the high-dimensional data problem, we employ deep learning techniques, which are well known for managing learning tasks with non-linear high-dimensional data such as language, images or video. However, as seen in the experimentation, and more specifically, in Fig. \ref{fig:ltf_preds_vs_targs}, the surroundings of the densities, which are constantly zero over time, can cause problems to the important parts of the phase space if they are not kept as zero during the propagation, creating artifacts in the main part of the distribution and eventually leading the forecast to an unrecognizable shape. Given that most of the phase space is in fact populated with zeros, and therefore, it has a sparse nature, our perspectives for future iterations of this work go through the use of graph-based representations. By representing each nonzero value of the phase space as a node in a graph, where adjacent cells of the phase space are edges, we can then apply a Graph Neural Network (GNN) that predicts how that graph will look like in future epochs (See Fig. \ref{fig:gnn}). This method learns directly from the original input space, taking advantage of the well-known scalability of GNNs for massive spatio-temporal data \cite{cini2022scalable}

Lastly, in relation to the final challenge previously highlighted, we propose evaluating MOCAT-ML using simulations of unexplored scenarios distinct from those utilized in the training set. This is done to gauge the generalizability of the model. All training and test simulations generated in this work by MOCAT-MC belong to one single scenario, namely one with an initial population and launch rate multiplied by a factor of $15$, where, at each time step within each Monte Carlo simulation, the number of objects within each two-dimensional bin of area-to-mass ratio and perigee is known. In order to proof the generalization capabilities of the model, this has to be tested on different scenarios to the ones used for training, to ensure that it learns the dynamics behind the evolution of the space environment, and not simply ``memorize" an optimal encoding to reproduce Monte Carlo simulations.

In addition, to objectively measure the model's performance relative to published results, we plan to reproduce two test cases examined by Giudici et al. in \cite{Giudici2023}, and pass them through our approach. Both cases involve fragmentation events of satellites from Starlink and OneWeb respectively, where the phase-space density distribution is propagated for up to one year following the fragmentation.

\section{Conclusions}
\label{sec:conclusions}
This study presents an advancement in the field of space environment modeling through the creation of MOCAT-ML, a machine learning-based framework designed to extends the capabilities of the MIT Orbital Capacity Tool (MOCAT) from a machine learning point of view. In this work, we focus on the predictive modeling of ASO density distributions in the Low Earth Orbit (LEO) environment. 

The innovative methodologies proposed in this research represent a breakthrough in addressing the computational efficiency challenge inherent in traditional space environment models. The experimental results demonstrate that some of the proposed deep learning architectures, particularly the end-to-end bi-dimensional autoencoder with ConvGRU cells, have a good performance in short-term forecasts, which they can deliver in the order of milliseconds. In contrast, they fail in forecasting densities up to more than two years ahead, for which several insights and ideas for improvement have been discussed.

The implications of this study are far-reaching. By improving the accuracy and reducing the computational demands of ASO density distribution modeling, MOCAT-ML offers a more efficient and effective tool for space traffic management and the long-term sustainability of the LEO environment. This is particularly crucial in the face of escalating challenges posed by the proliferation of satellites and space debris.

Moreover, our work underscores the potential of hybridizing Monte Carlo and machine learning predictions for space environment evolution. This perspective allows for initial epochs of propagation provided by traditional methods like MOCAT-MC, with subsequent predictions enhanced by the ML model, thereby offering a balanced strategy that maximizes accuracy and computational efficiency. We plan on exploring more in depth this integration of the two models in future iterations of the tool.

Future directions for this research also include refining the machine learning models to handle long-term forecasting horizons more effectively, testing the generalization capabilities of the model to unseen scenarios, quantifying prediction uncertainty, and exploring graph-based representations for the high-dimensional phase space data. By continuing to innovate in these areas, we aim to further advance our understanding, scalability and robustness of the applications of machine and deep learning in this field.



\section*{Acknowledgments}
This research was conducted during a fellowship at MIT funded by the Universidad Politécnica de Madrid’s program for promoting international academic collaboration in the USA, 2022/2023. The authors acknowledge the MIT SuperCloud and Lincoln Laboratory Supercomputing Center for providing (HPC, database, consultation) resources that have contributed to the research results reported within this paper/report. This work was partially supported by the National Science Foundation under award NSF-PHY-2028125.

\bibliography{sample}

\end{document}